\documentclass[twoside]{article}
\usepackage{qic,epsfig}

\begin{document}
\setlength{\textheight}{8.0truein}    

\runninghead{Integrated Optical Approach to  $\ldots$}
            {J. Kim and C. Kim}

\normalsize\textlineskip
\thispagestyle{empty}
\setcounter{page}{1}

\copyrightheading{0}{0}{2003}{000--000}

\vspace*{0.88truein}

\alphfootnote

\fpage{1}

\centerline{\bf
INTEGRATED OPTICAL APPROACH TO }
\vspace*{0.035truein}
\centerline{\bf TRAPPED ION QUANTUM COMPUTATION}
\vspace*{0.37truein}
\centerline{\footnotesize
JUNGSANG KIM\footnote{Electronic address: jungsang@ee.duke.edu}  and CHANGSOON KIM}
\vspace*{0.015truein}
\centerline{\footnotesize\it Electrical and Computer Engineering Department and Fitzpatrick Institute for Photonics}
\baselineskip=10pt
\centerline{\footnotesize\it Duke University, 130 Hudson Hall, Durham, NC 27708, USA
}
\vspace*{10pt}

\vspace*{0.21truein}

\abstracts{
Recent experimental progress in quantum information processing with trapped ions have demonstrated most of the fundamental elements required to realize a scalable quantum computer. The next set of challenges lie in realization of a large number of qubits and the means to prepare, manipulate and measure them, leading to error-protected qubits and fault tolerant architectures. The integration of qubits necessarily require integrated optical approach as most of these operations involve interaction with photons. In this paper, we discuss integrated optics technologies and concrete optical designs needed for the physical realization of scalable quantum computer.}{}{}

\vspace*{10pt}

\vspace*{3pt}
\communicate{to be filled by the Editorial}

\vspace*{1pt}\textlineskip    
\section{\label{sec:Introduction}Introduction}
 Quantum information processing with trapped ions has experienced a tremendous amount of progress in recent years since the initial proposal \cite{CiracPRL1995} and experimental demonstration \cite{MonroePRL1995}, both in improving critical experimental parameters for robust quantum computation and in architectural concepts leading to a more scalable construction of large-scale quantum information processors (QIPs). Critical milestones in the experimental progress include demonstration of robust quantum logic gates \cite{Schmidt-KalerNature2003,LeibfriedNature2003,HaljanPRL2005}, quantum teleportation \cite{RiebeNature2004,BarrettNature2004}, ion transport via electrostatic force \cite{RoweQIC2002}, long coherence times \cite{LangerPRL2005}, ion-photon entanglement \cite{BlinovNature2004}, remote ion entanglement \cite{MoehringNature2007}, quantum error correction \cite{ChiaveriniNature2004}, and multi-particle entanglement \cite{SackettNature2000,LeibfriedNature2005,HaffnerNature2005}. Based on these elements, architectures for realizing scalable quantum computation on an integrated ion trap chip have been proposed \cite{KielpinskiNature2002}. Further improvements on these basic ideas have been pursued both from technological \cite{KimQIC2005} and computer architecture perspectives \cite{MetodiMICRO2005}. These experimental and architectural developments led to studies of on-chip planar ion traps \cite{PearsonPRA2006,SeidelinPRL2006,Brown0603142}, as a first step toward an integrated processor. It is apparent that the task of assembling a large-scale quantum information processor based on ion traps is largely a technology challenge, resembling the integrated circuits (IC) technology for classical processors based on silicon.

The implementation of a scalable ion trap QIP requires a scheme that allows integration of {\em all necessary functionalities}, which include qubit preparation, manipulation, communication and detection. The necessary hardware components consist of an integrated ion trap chip, controllers to create and distribute the voltage waveforms needed for ion transport, laser systems that create the required control beams, a beam delivery network to distribute  laser beams to multiple ion locations, a scalable state detection strategy, and a classical real-time controller to manage all control elements \cite{KimQIC2005,Steane0412165}. Approaches to interconnect multiple ion trap chips with a network of quantum communication channels are essential components for scalable architecture. In this paper, we summarize the optical requirements for preparation, manipulation, detection and communication of trapped ion qubits and outline integrated approaches to providing these optical functionalities. While it is technologically premature to design and realize a realistic QIP at this stage, the approaches considered in this paper provides substantial advantage over conventional experimental techniques in terms of performance and scalability.

We start out in Section \ref{sec:ArchitecturalVision} with an overall architecture for constructing a scalable QIP. In Section \ref{sec:PhysicalRequirements}, we summarize the optical requirements and system design considerations for maximally utilizing the limited resources in the QIP system. Based on these considerations, we discuss a possible layout for the required laser beams in Section \ref{sec:BeamDesign} for quantum logic operation, cooling and state detection. Section \ref{sec:Measurement} discusses the issues related with detection of multiple ion qubit states. Section \ref{sec:Communication} considers optical designs for realizing  entangled ion pair via photon exchange. The conclusions are presented in Section \ref{sec:Conclusions}.

\section{\label{sec:ArchitecturalVision} Scalable architecture for ion trap QIP}
\begin{figure}[htbp]
	\centering
		\includegraphics[width=4in]{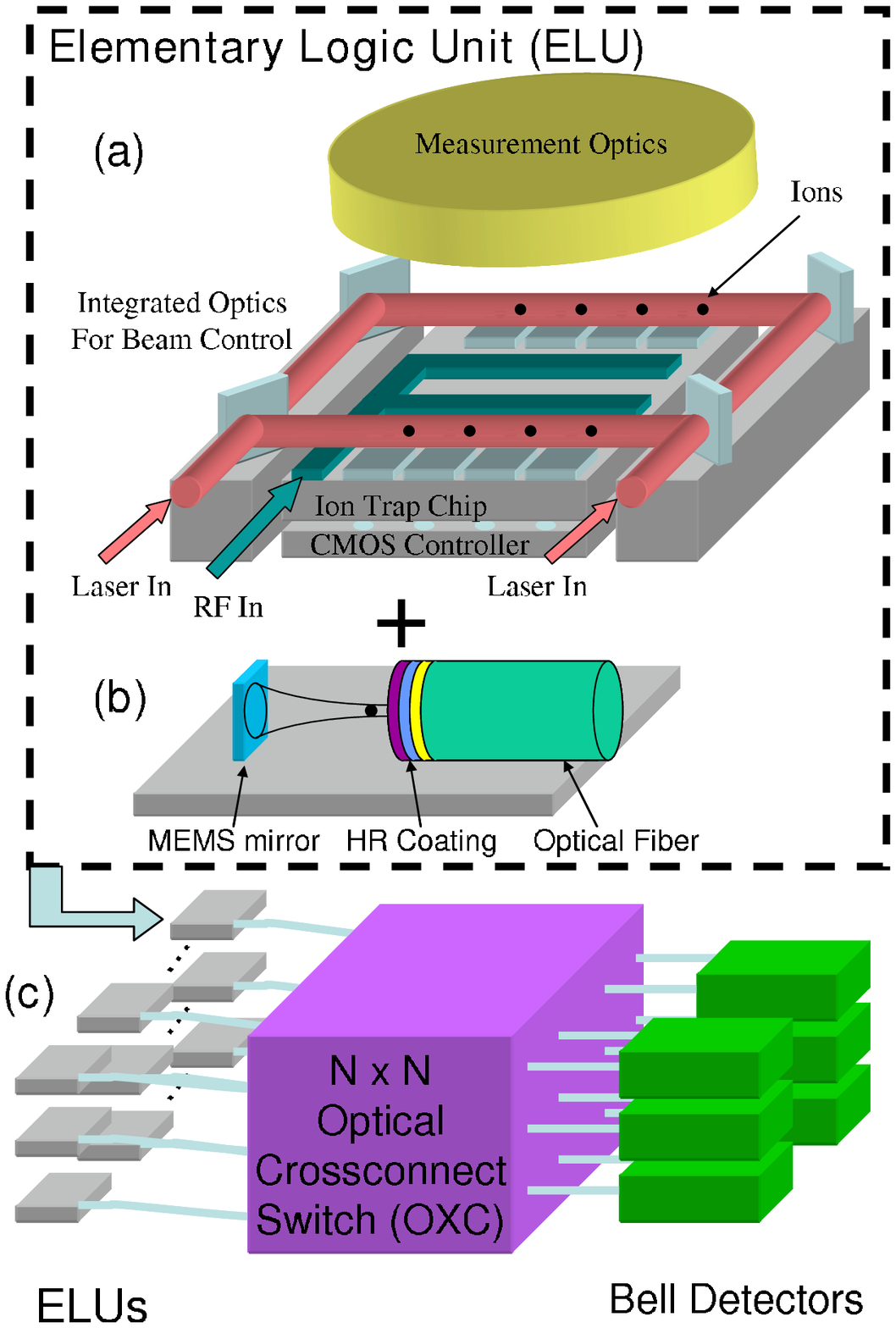}
		\fcaption{\label{EntanglementGeneration}
(a) Schematic of integrated ion trap QIP on a chip, as outlined in Ref. \cite{KimQIC2005} (Reprinted with permission). (b) Schematic setup  to increase the collection efficiency of spontaneously emitted photons from a single ion using a micro-cavity on the chip. The combination of the two features described in (a) and (b) forms an elementary logic unit (ELU) for the scalable QIP. (c) Schematic of creating entanglement between arbitrary pairs of ELUs through an optical crossconnect switch \cite{MoehringJOSAB2007}.}
\end{figure}

Scalable architectures for ions integrated onto a chip have been proposed and studied in recent years \cite{KielpinskiNature2002,KimQIC2005}. The first goal in scalable QIP is to create a technology platform that will enable a chip with hundreds of trapped ions, on which flexible manipulation of ion qubits can be implemented. Figure \ref{EntanglementGeneration}a shows a schematic of an approach outlined in Ref. \cite{KimQIC2005}. The planar ion trap chip technology is under intense development \cite{IonTrapWorkshop2006}, while the integration of the optical elements still remains a major challenge. We discuss the constraints and strategies for integrating the optical control signal and detection elements in the next four sections.

A chip that integrates hundreds of ions can represent only a handful of {\em logical} qubits when quantum error correction codes are employed to implement a fault-tolerant architecture. In order to represent a much larger number ({\em e.g.}, 100) of logical qubits, the total number of physical qubits needs to increase by about two orders of magnitude. The physical size of the ion trap chip will be limited by, among other things, the diffraction of laser beams that can be brought onto the chip (see Section \ref{sec:BeamDesign}). Furthermore, all useful quantum algorithms require a global entanglement among all participating (logical) qubits, in such a way that the problem cannot be segmented into smaller blocks. This implies that the qubits in the computer needs to be transported globally, and the number of physical ions that can be trapped on a single chip will be limited by the complexity of the interconnect architecture on the chip as the number grows \cite{MetodiMICRO2005}. A process that probabilistically generates  entanglement between two remote quantum nodes can be an important architectural element in overcoming these limitations to construct a scalable quantum information processor \cite{DuanNature2001,DuanQIC2004,DuanPRA2006,OiPRA2006}. Once remote entanglement is generated, quantum logic operations can be performed by teleportation gates \cite{GottesmanNature1999}. Recent preliminary studies on the issue of qubit distribution within the QIP using teleportation \cite{IsailovicISCA2006,vanMeterISCA2006} highlight benefits of widely adopting the scheme for scalable QIP construction. In order to enable such an architecture, each ion trap chip must be equipped with a communication port (which we will call a ``quantum teleport'') where the spontaneous emission from a single ion can be efficiently captured into a single mode fiber, as shown in Fig. \ref{EntanglementGeneration}b. We identify a chip with a quantum teleport as an elementary logic unit (ELU) that represents a few logical qubits. Utilizing a transparent optical crossconnect switch \cite{KimPTL2003,NeilsonPTL2003}, a flexible quantum network can be formed to create entanglement between arbitrary pairs of ELUs in the QIP architecture (Fig. \ref{EntanglementGeneration}c) \cite{MoehringJOSAB2007}. The use of transparent optical switches as quantum channels of communication has been demonstrated in the context of quantum cryptography \cite{ToliverPTL2003}. This approach enables entanglement generation process where the cost is independent of the distance between the qubits, which can be utilized to implement an ``abstract concurrent'' architecture for the QIP \cite{vanMeterPRA2005} shown to provide a substantial improvement in computation time by employing efficient algorithms for basic arithmetic operations \cite{DraperQuantuPh0406142}. The architectural advantage of this strategy will critically depend on the performance of the hardware used to create the entangled ion pairs. In Section \ref{sec:Communication}, we will discuss the hardware requirements for realizing an effective quantum teleport.

\section{\label{sec:PhysicalRequirements}Optical requirements and design strategy}
\subsection{\label{subsec:OpticalRequirements} Choice of qubits and laser requirements}
\begin{figure}[htbp]
	\centering
		\includegraphics[width=4in]{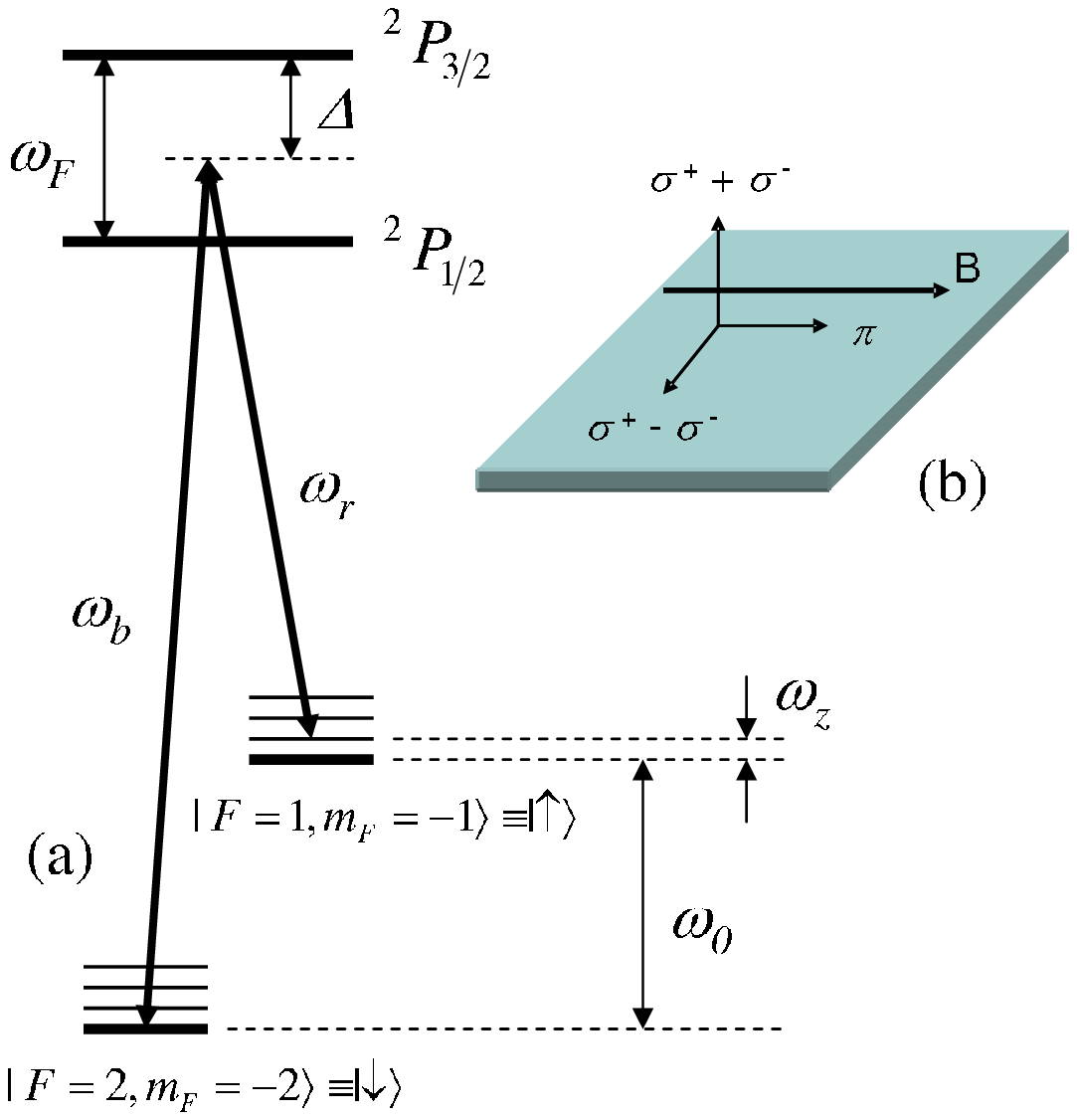}
		\fcaption{\label{EnergyLevels}
(a) Relevant energy levels for the $^{9}$Be$^{+}$ ion. $\omega_0$, $\omega_F$ and $\omega_z$ denote the ground state hyperfine splitting, upper state fine splitting and motional energy level, respectively. $\Delta$ is the detuning from the resonant transition, and $\omega_b$ and $\omega_r$ denote the blue- and red-detuned Raman laser beams, respectively.  (b) Definition of the polarization directions with respect to the magnetic field direction.}
\end{figure}

The wavelength, intensity, polarization and propagation direction requirements for the laser beams are determined by the choice of ion species, type of cooling scheme used, and type of gate operations employed. We concentrate our analysis on $^{9}$Be$^{+}$ as qubit ions and $^{24}$Mg$^+$ as ions used in sympathetic cooling \cite{BarrettPRA2003}, in a planar trap above the surface of the trap substrate \cite{KimQIC2005,SeidelinPRL2006,Brown0603142}. The qubit states are two hyperfine ground states of  $^{9}$Be$^{+}$ ion ($|F=2,m_F=-2\rangle=|\downarrow \rangle$, $|F=1,m_F=-1\rangle = |\downarrow\rangle$), and we consider the geometric phase gate as the two-qubit gate of choice \cite{LeibfriedNature2003}. The schematic of relevant energy levels of the $^{9}$Be$^{+}$ ion is shown in Fig. \ref{EnergyLevels}a. We refer the details of the ion trap physics to other references \cite{WinelandJRNIST1998,KingThesis1999,WinelandPTRSLA2003}. Here, we summarize the polarization and intensity requirements for laser beams necessary for various qubit operations at a high level relevant for integrated optical system design. The intensity requirements arise from the target fidelity of the single- and two-qubit gates \cite{OzeriPRA2007}, and are described in three categories for simplicity: mild ($\sim$ few $\mu$W), modest ($\sim$ few mW) and extreme ($>$ 100 mW). The polarization of the laser beams is defined with respect to the quantization axis determined by an external magnetic field ($\vec{B}$) applied to the ions. The direction of the magnetic field and various polarization notations ($\pi$, $\sigma^+ \pm \sigma^-$) are shown in Fig. \ref{EnergyLevels}b.

\begin{table}[hb]
\tcaption{\label{Table1} Summary of requirements for various beams needed in the QIP under consideration.}
\centerline{\footnotesize\smalllineskip
\begin{tabular}{ccccccc}
 \hline
                      & & Target & Raman & Momentum && \\
 Function	&	Polarization & Ion & Detuning & Difference & Intensity & Location \\
 \hline
 &&&&&& \\
 RSRC & $\pi$, $\sigma^+$ or $\sigma^-$ & $^{24}$Mg$^+$ & $\omega_0 ^\prime - \omega_z$\footnote{$\omega_0 ^\prime$ is hyperfine ground state splitting of $^{24}$Mg$^+$.} & Large $\Delta k$ & Modest & All Gate Regions \\
 Re-pumping & $\sigma^+$ or $\sigma^-$ & $^{24}$Mg$^+$ & - & - & Mild & All Gate Regions \\
 Single qubit	& $\pi$, $\sigma^+$ or $\sigma^-$ & $^{9}$Be$^{+}$ & $\omega_0$ & Small $\Delta k$ & Modest & Single Qubit Gate Regions \\
 Two qubit	& $\sigma^+$+$\sigma^-$, $\sigma^+$-$\sigma^-$ & $^{9}$Be$^{+}$ & $\sqrt{3}\omega_z + \delta$ & Large $\Delta k$ & Extreme & Two Qubit Gate Regions \\
 Measurement	& $\sigma^-$ & $^{9}$Be$^{+}$ & - & - & Modest & Measurement Regions \\
 Doppler	& $\sigma^-$ & $^{9}$Be$^{+}$ & - & - & Mild & $^{9}$Be$^{+}$ Loading Zone,\\
 &&&&&& Measurement Regions \\
 Depopulation	& $\sigma^-$ & $^{9}$Be$^{+}$ & - & - & Mild & $^{9}$Be$^{+}$ Loading Zone,\\
 &&&&&& Measurement Regions \\
 Doppler	& Any & $^{24}$Mg$^+$ & - & - & Mild & $^{24}$Mg$^+$ Loading Zone \\
\end{tabular}}
\end{table}

The primary cooling mechanism used in the logic gate regions is the resolved-sideband Raman cooling (RSRC) process on $^{24}$Mg$^+$ ions, which requires two Raman beams and one re-pumping beam \cite{MonroePRL1995b}. The single qubit and two-qubit operation requires two Raman beams to be applied to the qubit ions. Measurement beams, Doppler cooling beams and depopulation beams (required in case spontaneous emission takes the qubit ions outside the qubit states) are close-to-resonant beams needed in the measurement regions and the initial loading zones. Table \ref{Table1} summarizes the requirements for various beams.

\subsection{\label{subsec:SystemConsiderations} System design considerations}
Overall layout of the QIP will depend on higher level system considerations such as the choice of error correcting codes and fault-tolerance schemes. Design considerations for such higher level architectures are in its infancy \cite{MetodiMICRO2005}, and their relevance is heavily constrained by the availability of hardware capabilities at the physical layer. We first consider system design considerations within an ELU (Fig. \ref{EntanglementGeneration}a).

\begin{figure*}[htbp]
	\centering
		\includegraphics[width=5.6in]{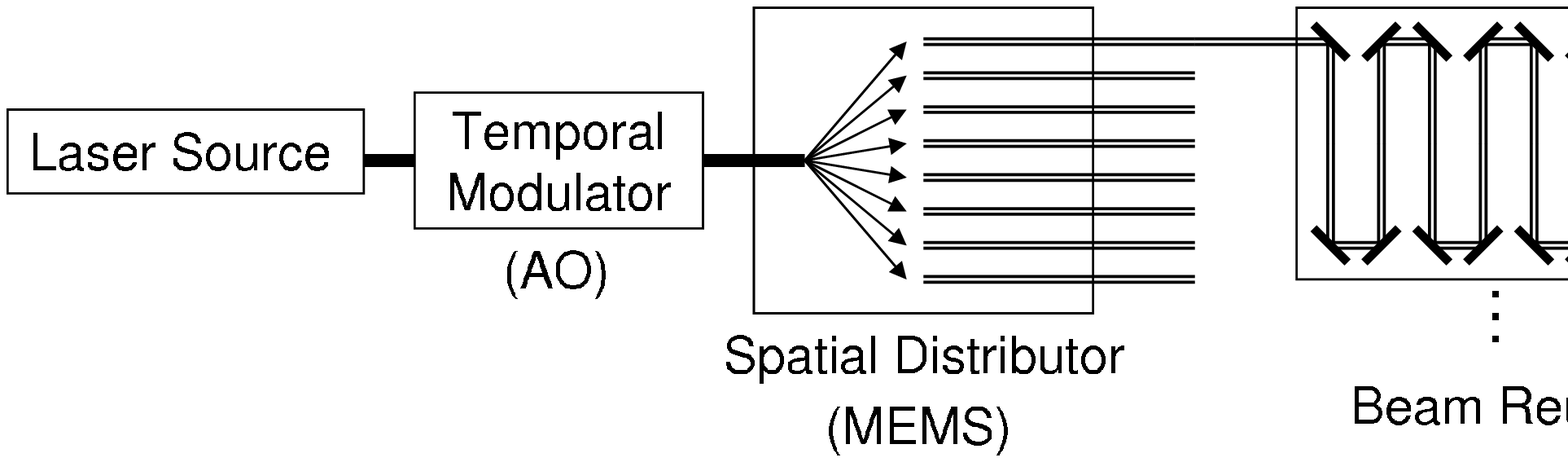}
		\fcaption{\label{BeamDistribution}
Strategy for beam distribution for the QIP. The laser source is prepared to provide necessary stability, and then modulated temporally using a fast modulator [typically an acousto-optic (AO) modulator]. The modulated beam is distributed using an optical system capable of directing the beam to several spatial locations (using MEMS technology). The resulting beam can be re-used for multiple operations if appropriate before being dumped.}
\end{figure*}

The laser beams necessary for the logic gate operations need to be stabilized in terms of frequency, intensity and polarization. Errors arising from the spontaneous emission process during the gate operation calls for higher power in the Raman lasers \cite{OzeriPRA2007}, so the laser power is one of the most expensive resources in the QIP construction. Figure \ref{BeamDistribution} shows the strategy for effective utilization of the laser beams in the QIP. The beam from a stabilized laser source is tailored in two stages: first it is modulated using a fast temporal modulator, and then distributed to multiple locations on the QIP using an optical multiplexer based on a flexible beam steering system. This approach decouples the requirements for high speed temporal modulation and flexible spatial distribution, where optimum technology can be chosen to achieve each functionality ({\em e.g.,} acousto-optic or electro-optic modulator for fast temporal modulation and MEMS technology for spatial distribution). After spatial distribution, further integrated optics technologies can be utilized to re-use the beam over multiple ions (or ion pairs) whenever feasible. Such scheme can prove useful if one can utilize single-instruction on multiple-data (SIMD) architectures for the QIP design \cite{KimQIC2005,WhitneyNSC2003}.

Besides the temporal modulation and spatial distribution of the laser beams, ion transport can be used to select the ions on which the logic gates are operated on. Efficient QIP operation requires optimal arrangement of these three control mechanisms to achieve maximum useful logic operations per unit time. The time constants for temporal modulation of the optical beams, transporting ions to the proper zones for logic gate operation, and re-configuration of the beam paths must be carefully considered for the optimized hardware architecture. Other time constants affecting the QIP operations are qubit measurement times and entangled pair generation times. Advanced concepts of scheduling, pipelining and SIMD principles heavily used in computer architectures can be applied to the QIP hardware design \cite{PattersonHennessy2005}.

\section{\label{sec:BeamDesign} Optical design for beam configuration}
\begin{figure}[htbp]
	\centering
		\includegraphics[width=5in]{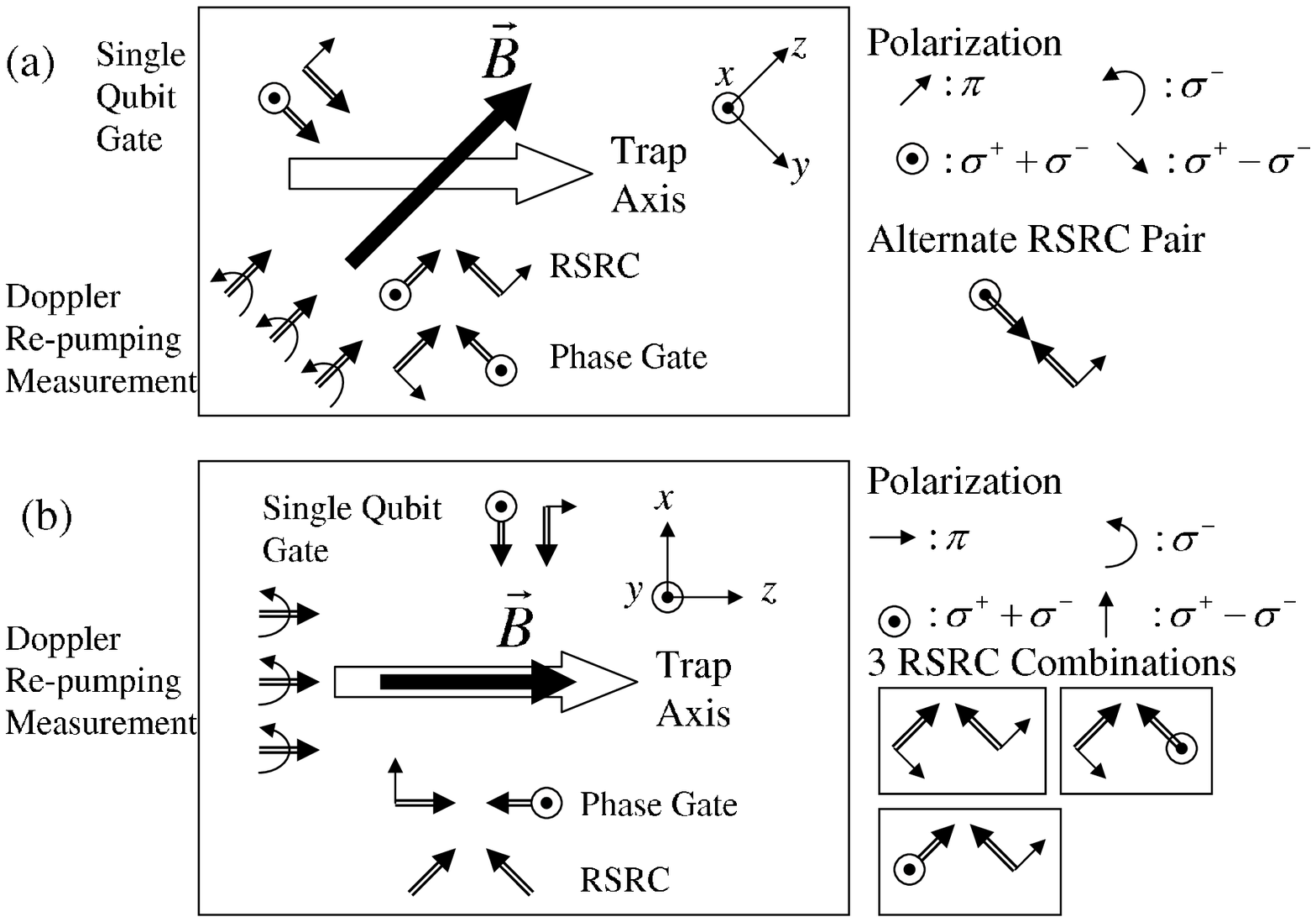}
		\fcaption{\label{BeamArrangement}
Beam arrangement schemes for the QIP, depending on the alignment between the trap axis (open arrow) and the magnetic field $\vec{B}$ (dark arrow). The double arrows indicate the propagation direction of the beams, while the small arrows indicate the polarization of the electric field. (a) Scheme A, where $\vec{B}$ field is applied 45$^\circ$ with respect to the trap axis. In this case, the RSRC beams can be 90$^\circ$ with each other, or counter-propagating. (b) Scheme B, where $\vec{B}$ field is parallel to the trap axis. In this case, there are three possible polarization combinations for the RSRC beams. }
\end{figure}

In this section we consider possible schemes for arranging the beams on the QIP chip. Since the detection beam requires pure circular polarization, this beam must propagate parallel to the magnetic field $\vec{B}$, which  forces the magnetic field to be parallel to the chip surface. The trap axis also lies in the plane of the chip surface (deviations can be induced for Doppler cooling purposes where necessary \cite{SlusherPrivate}). In order to accommodate all the necessary beams described in Table \ref{Table1}, we find two possible alignments between the trap axis and the $\vec{B}$ field: Scheme A, where they make a 45$^\circ$ angle (Fig. \ref{BeamArrangement}a) and Scheme B, where they are aligned parallel to each other (Fig. \ref{BeamArrangement}b). While these beam arrangements satisfy the necessary condition for QIP operation, further design optimization is desired to maximize the system operation within the resource constraints as discussed in Sec. \ref{subsec:SystemConsiderations}. Since the two-qubit phase gate beams require the largest optical power, we concentrate on the optical design that most efficiently utilizes photons in these beams.

\begin{figure}[htbp]
	\centering
		\includegraphics[width=4in]{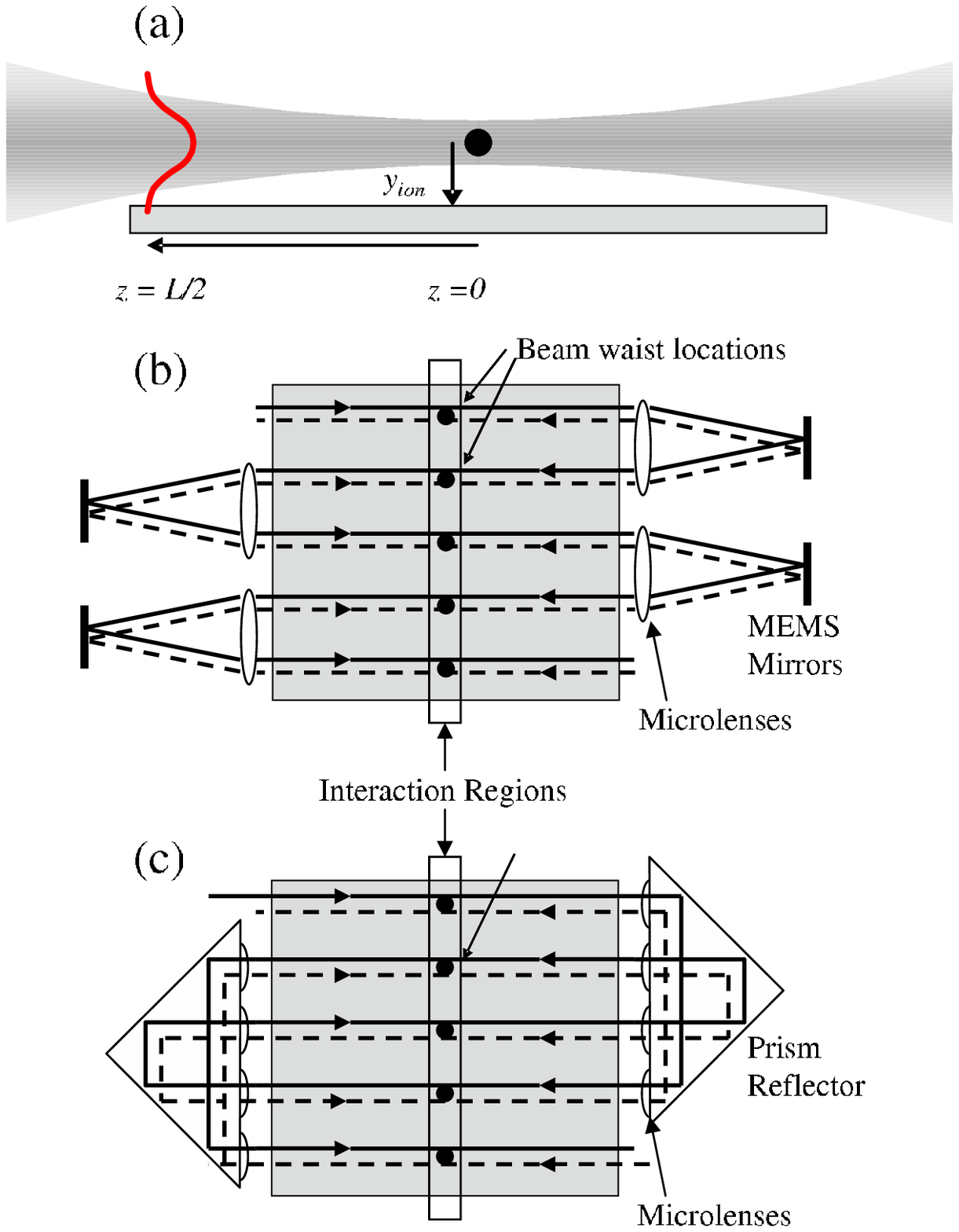}
		\fcaption{\label{BeamDetails}
(a) Gaussian beam aligned to address the ion trapped above the surface of the trap chip (side view). (b) Beam recycling strategy for counter-propagating Raman beams using MEMS mirrors. Focusing microlenses and MEMS mirrors image the Gaussian beam waist to the next ion location. (c) Beam recycling strategy using prism reflectors. The microlenses are integrated onto the prism surface for imaging the Gaussian beam waist. The counter-propagating beams in (b) and (c) are shifted for clarity.}
\end{figure}

One needs to find the optimum design for the Gaussian beams addressing the ions. The optimum Gaussian beam waist is determined by the size of the ion trap chip (linear dimension of $L$) and the height of the ions above the chip surface ($y_{ion}$), as shown in Fig. \ref{BeamDetails}a. The $1/e^2$ radius of the Gaussian beam at the edge of the chip must be minimized to reduce photon scattering from the surface of the chip, which induces undesired Stark shifts and gate errors. Assuming the ion is located at the center of the chip, the beam radius at the edge of the chip
\begin{equation}
W(L/2) \, = \, W_0 \left[1\, + \, \left(\frac{L\lambda}{2 \pi W_0^2}\right)^2\right]^{1/2}
\label{Equation1}
\end{equation} 
is minimized at the optimum beam waist $W_{0,opt} = \sqrt{L\lambda / 2 \pi}$. For $L$=10mm and $\lambda$=313nm, the optimum beam waist is $W_{0,opt} \simeq 22.3 \mu$m, and $W(L/2=5$mm$)=31.6\mu$m. Therefore, if the ions are trapped $\sim 50\mu$m above the surface, the fraction of photons out of the laser beam hitting the chip surface will be less than 0.08\% on each side of the Gaussian beam. It decreases rapidly as the size of the chip $L$ is reduced, or as the trap height $y_{ion}$ is increased. This consideration also puts a practical limit on the size of the ion trap chip to be less than about 1 cm on a side.

The RSRC, single qubit gates and phase gates involve stimulated Raman transitions between the ground states of the ion. The Raman transition matrix elements depend on the amplitude and polarization of the optical fields \cite{WinelandJRNIST1998,WinelandPTRSLA2003}. Since the Raman transition adds and subtracts only a single photon from each classical control beam, the coherent optical fields driving the Raman transition remain virtually unchanged (and disentangled) from the internal degree of freedom for the ion. This opens up the possibility of recycling the Raman beams for identical logic operations on other ions or ion pairs. Recycling the Raman beams is feasible from the hardware perspective only if the interaction Hamiltonian for the fields and the ions (or ion pairs) of interest remains identical. It is only meaningful from the applications perspective if a higher level QIP architecture calls for the need to apply identical logic gates to multiple ions (or ion pairs). Transversal implementation of fault-tolerant logic gates on encoded qubits would benefit from such parallel arrangement \cite{NielsenChuang2000}.

One convenient scheme to realize this is if the two Raman beams are counter-propagating with respect to each other, as shown in Fig. \ref{BeamDetails}b and c. In Fig. \ref{BeamDetails}b, the distance between an interaction region (where a Gaussian beam waist is located) and a microlens, and the distance between the microlens and a MEMS mirror are equal to the focal length $f$ of the microlens. In this arrangement, the beams exciting an interaction region are reflected and re-focused onto the next interaction region to address the next ion (or ion pair). Another way to achieve this successive imaging is to use the total internal reflection in a prism, as shown in Fig. \ref{BeamDetails}c. For this arrangement, the size of the prism, the distance between the interaction region and the prism, and the focal length of the lens must be chosen to allow exact imaging of the beam waist from one interaction region to the next. The total number of interaction regions covered is determined by the relative shift of the two prisms with respect to each other. An example arrangement to use the beams for 5 interaction regions is as follows: for a  chip with width $L$, one can fabricate microlenses with focal length $f=L/2$ on the hypotenuse of the prisms. Two prisms with side $L/\sqrt{2}$ (hypotenuse $L$) have four microlenses with centers located at 0.2$L$, 0.4$L$, 0.6$L$ and 0.8$L$ are aligned on either side of the chip with a relative offset of 0.2$L$. Then, five beam waists will be located in the middle of the chip, separated by 0.2$L$, as shown in Fig. \ref{BeamDetails}c.

The lenses and mirrors (or the prisms) have finite optical loss. One important advantage of these optical designs is that the two beams traversing the paths experience exactly the same optical loss per reflection, provided that (1) the optical components are ideal and (2) the reflectivity and loss are independent of polarization and laser frequency. If the electric field amplitudes of the two Raman beams entering the system are denoted by $E_r$ and $E_b$, and the reflection coefficient due to each optical retro-reflection (including the loss due to lenses and the prism) is characterized by $r_i$ ($i=1,2,\ldots,N$, where $N$ is the total number of optical reflections), the product of the two field amplitudes at any given interaction region is given by $E_r E_b \prod_{i=1}^N r_i$. Since the Rabi frequency of the Raman transition only depends on the product of the two fields to first-order, these schemes can be utilized to perform the SIMD operation on multiple qubit pairs using a single pair of beams.

It is important to note that there are practical challenges in realizing these schemes. While the product of the two field intensities is constant across the interaction regions, the field strength of each beam is not identical. This discrepancy can potentially lead to differences in the AC Stark shift for the ions at different locations. For the geometric phase gate considered here, the impact of the AC Stark shift on the gate fidelity can be eliminated by careful choice of the polarizations of the Raman beams \cite{WinelandPTRSLA2003}. This is not the case in most other gates, but methods for reducing the sensitivity to AC Stark shifts for other two-qubit schemes are under consideration \cite{WesenbergNIST2006,ChuangPrivate}. Another challenge arises from the precision to which the Rabi frequency of the Raman transitions have to be maintained. In real experiments, the fidelity of the gates is maximized by carefully tuning the duration of the Raman beams that interact with the ions, to account for any intensity modifications that cannot be compensated for by the optical components. The product of the two electric field amplitudes must be maintained to within a percent among all interaction zones, in order to maintain the fidelity of the gate operation. One possible method is to control the location of the ions along the length of the beam using segmented electrodes, to fine-tune the Rabi frequency in the presence of component variations. Lastly, unwanted micromotion of the ions that modulate the laser field experienced by ion must be minimized. This can be achieved by carefully controlling the compensation voltages on DC electrodes on the chip \cite{SlusherPrivate}.

\begin{figure}[htbp]
	\centering
		\includegraphics[width=5.6in]{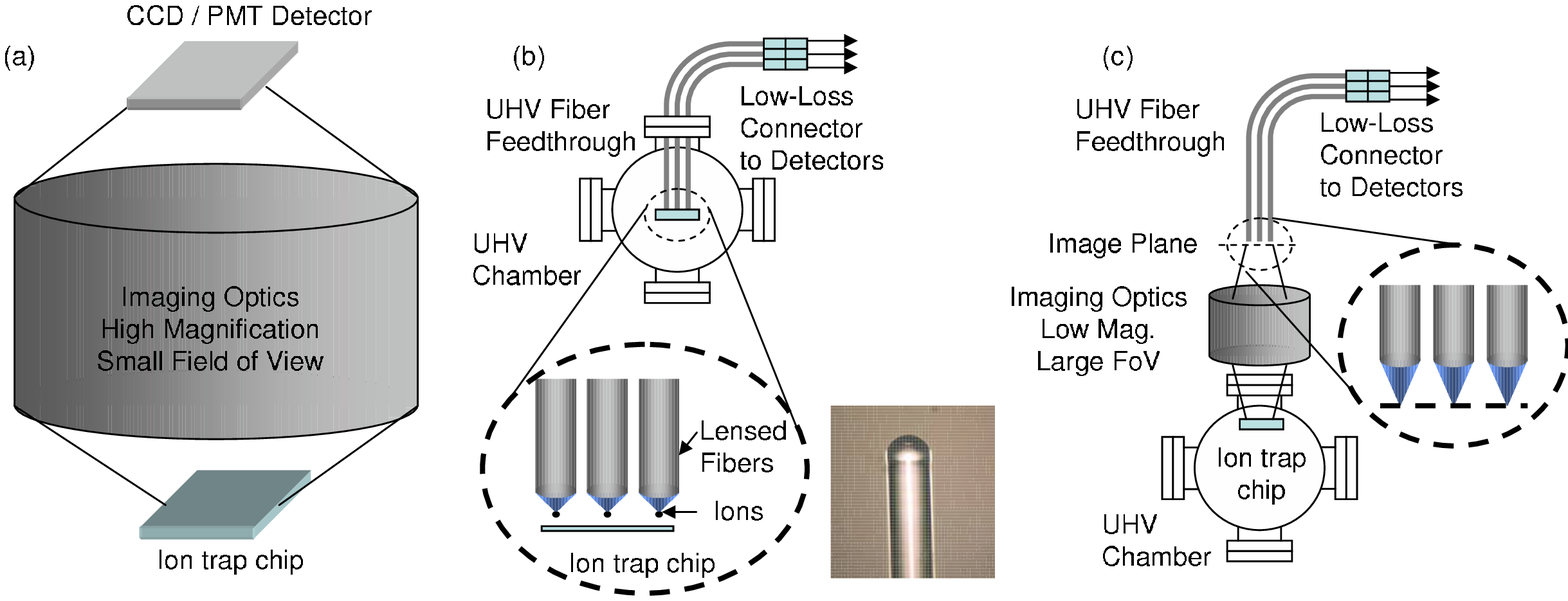}
		\fcaption{\label{DetectionStrategy}
(a) Traditional approach to detection of photons scattered from the ion detection process. (b) Integrated and scalable approach using micro-optical components located close to the trapped ions and arrayed detectors. The inset shows an example of a micro-lensed fiber fabricated by melting the fiber tip, featuring radius of curvature of about 230 $\mu$m. (c) A hybrid solution of using a low magnification imaging optics with micro-optical components. This approach avoids feeding fiber arrays into the UHV chamber.}
\end{figure}

\section{\label{sec:Measurement} Integrated approach to ion state detection}
A state dependent cycling transition process is utilized for the qubit measurement, where only one qubit state scatters photons when illuminated with a laser beam \cite{NagourneyPRL1986,BergquistPRL1986,SauterPRL1986}. The rate of scattered photons during the state measurement process is determined by the spontaneous emission lifetime of the excited states used for cycling transition. For $^9$Be$^+$ ions, the natural linewidth of the upper state corresponding to spontaneous emission is $\sim 20$MHz, corresponding to a maximum scattering rate of about $6 \times 10^7$ photons per second (similar for other ions) \cite{OzeriPRA2007}. Ideal collection optics with $F/\# \sim 1$ can collect about 5$\%$ of these photons: we will assume that typical collection optics directs about $10^6$ photons per second towards the detector \cite{LeibfriedPrivate}. Current experimental setups utilize custom fabricated optical lenses located outside the UHV chamber that achieve low $F/\#$ collection. The collected photons are detected using photomultiplier tubes (PMTs) or high efficiency charge-coupled device (CCD) cameras for a measurement duration long enough to distinguish the photon scattering state (Fig. \ref{DetectionStrategy}a). The signal-to-noise ratio achievable in current experiments requires measurement times to be $\sim 100 \mu$s, and poses a time-limiting step in the computation process. Furthermore, this method is not scalable to a large number of detection zones due to bulky optical elements and detectors.

\subsection{\label{subsec:Detector} Choice of photon detectors}
In order to understand the system requirements for effective state detection, we consider a photon flux $P$ ($\sim 10^6$ photons/sec) arriving at the detector with efficiency  $\eta$, which includes both the quantum efficiency (QE) of the detector and the collection efficiency of the optics between the low $F/\#$ collection lens and the detector. A primary photocurrent $I_p = e \eta P$ (A) is generated at the detector ($e$ is the electron charge), which has an internal gain process characterized by the gain $M$ and the excess noise factor (ENF) $F$ \cite{KimBook2001}. The resulting (amplified) current or charge (accumulated for a finite integration time) is amplified by electronics that feature an input current noise of $i_{th}^2$ (A$^2$/Hz). The detector system also suffers from shot noise arising from the random arrival and detection of photons, and a dark (background) current $I_B$ (or, dark counts per unit time) intrinsic to the detector or arising from the scattered pump laser. In order to minimize the impact of the dark current, (1) the detector area should match the ``size'' of each ion imaged onto the detector and (2) this area should be minimized while the detector operates below saturation. Under these assumptions, the figure of merit one can use is the bit-error-rate (BER) of distinguishing the photon scattering state in the presence of detector/readout noise. One can calculate the number of signal electrons $S$ collected and noise-equivalent electrons $N$ generated during the measurement time interval $T_M$ as
\begin{eqnarray}
e^2S^2 & = &  (I_p  M T_M)^2 \\
e^2N^2 & = & \left[(i_{th}^2 + 2 e I_p M^2 F) B+I_B^2 M^2 \right] T_M^2,
\label{Equation2}
\end{eqnarray} 
where $B = 1/T_M$ is the measurement bandwidth. In a typical readout circuit, a transimpedance amplifier (TIA) is used as the first-stage amplifier to convert the (amplified) photocurrent into a voltage signal. The input current noise $i_{th}^2$ arises from the thermal noise of the input load impedance and the intrinsic noise of the amplifier element. The thermal noise contribution dominates over intrinsic amplifier noise when the transimpedance gain is large, which is a necessary condition for low light level detection. When the measurement bandwidth $B$ and the capacitance of the device $C$ is given, one can choose the load impedance of the TIA to be $R_L = 1/CB$. In this optimum case, the amplifier noise contribution is minimized with a lower bound of $i_{th}^2 B T_M^2= 4k_B T C$ independent of the integration time, where $T$ is the operating temperature of the circuit. For detectors with an internal gain, the thermal noise of the amplifier becomes negligible compared to the shot noise of the input signal if the gain is large enough. For example, the signal-to-noise ratio (SNR) is purely limited by the shot noise of the photon detection events when $M \geq \sqrt{4k_B TC}/I_p T_M \simeq 800/P$ at $T=$ 300K and $C=$1 pF, where $P = I_p T_M/e$ is the total number of photons absorbed by the detector during the integration time (for typical measurement, $P \geq 5$). In the long integration time limit, the SNR is eventually limited by the background current (or background counts) intrinsic to the detector.

In this analysis, we consider five different detectors with the following assumptions \cite{KimBook2001}.
\begin{itemlist}
\item Ultra-violet photon counter (UVPC), a hypothetical device with operating characteristics very similar to visible light photon counter (VLPC) with an enhanced response in the UV wavelength range \cite{AtacNIMPR1992,KimAPL1997,KimAPL1999,TakeuchiAPL1999}. We assume the UVPC has a quantum efficiency (QE) of $\eta_Q \simeq 65 \%$, a gain of $M = 3 \times 10^4$, an ENF $F=1.0$ and a dark count rate of $\sim 20,000$ counts per second (cps).
\item Photomultiplier tube (PMT) that has QE of $\eta_Q = 10\%$, gain of $M = 10^6$, ENF $F=1.5$ and dark count rate of $\sim 500$ cps.
\item Charge-coupled device (CCD) camera, with a QE of $\eta_Q = 65\%$, a gain of $M = 1$, ENF $F=1$ and a dark count rate of $\sim 100$ cps. The CCD provides a very large number of pixels ($\sim 10^6$), and the pixels are read out sequentially using several channels of readout circuits each running at around 10 MHz. The frame rates for reading out all the pixels are rather slow ($60-15,000$ frames per second). The higher end of frame rates can be achieved by effectively reducing the number of pixels in the camera frame.
\item Electron multiplying CCD camera (EMCCD), with a QE of $\eta_Q = 65\%$, a gain of $M = 100$, an ENF $F=2$ and a dark count rate of $\sim 100$ cps. In such a camera, the readout electronics consist of a chain of gain stages that feature a small gain (1.004-1.015) per stage. Cascading a large number (400-600) of such gain stages leads to large gains (100-1,000) with an ENF close to 2 \cite{RobbinsIEEEED2003}. While these provide close-to-ideal gain characteristics, there is a net latency (of $40-60 \mu$s) in the readout of each pixel due to the execution time of the gain element.
\item Avalanche photodiodes, with a QE of $\eta_Q = 50\%$, a gain of $M = 100$, an ENF $F=10$ and a dark count rate of $\sim 100$ cps \cite{GramschNIMPR1999}.
\end{itemlist}

\begin{figure}[htbp]
	\centering
		\includegraphics[width=4.5in]{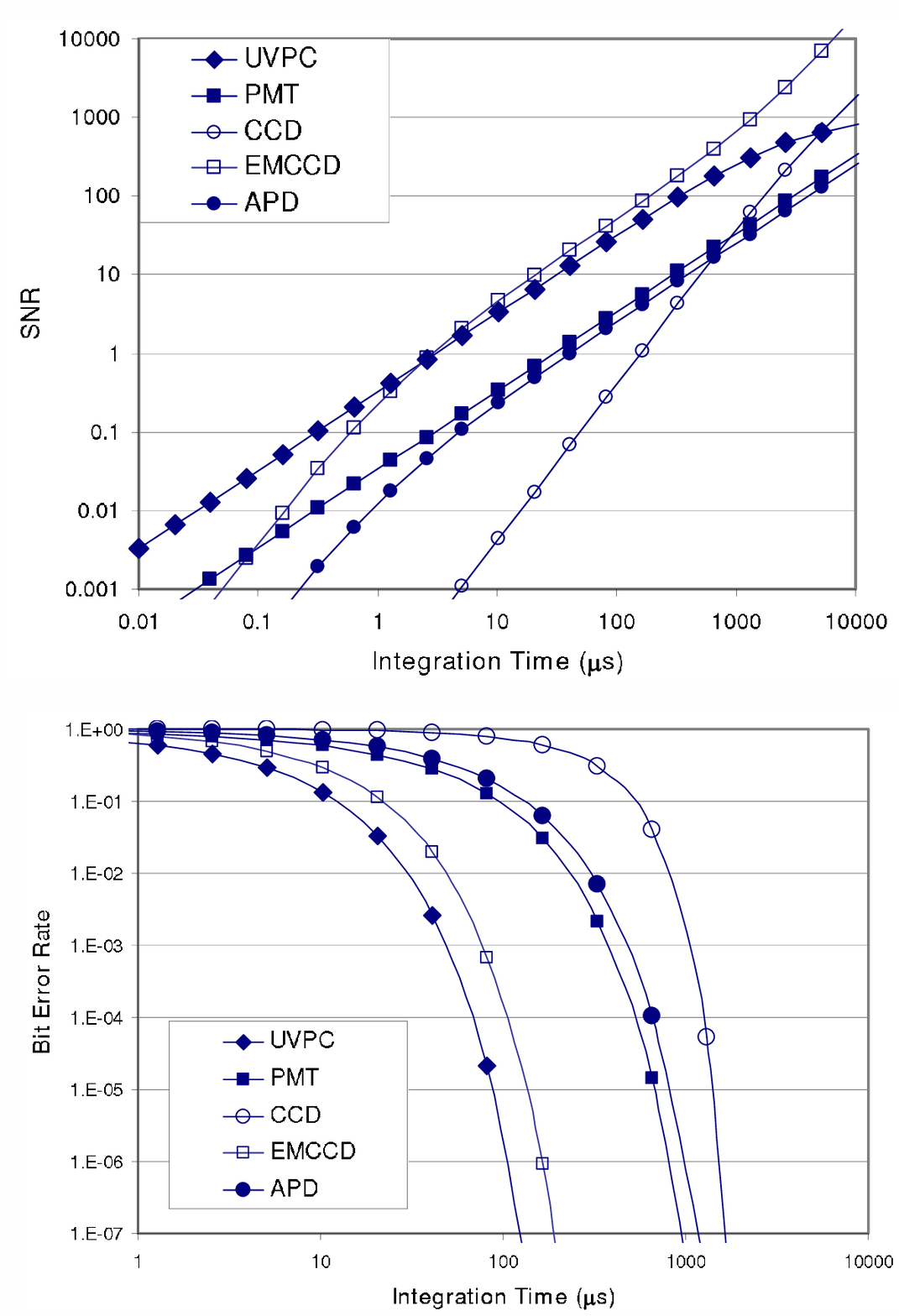}
		\fcaption{\label{SNRAnalysis}
(a) Signal-to-noise ratio (SNR) and (b) bit-error-rate (BER) as a function of integration time for various detector devices, based on the assumptions outlined in Sec. \ref{subsec:Detector}. The devices considered are UVPC (solid diamonds), PMT (solid squares), CCD (open circles), EMCCD (open squares) and APD (solid circles). In (b), the threshold for each operating condition is optimized to minimize the BER.}
\end{figure}

Figure \ref{SNRAnalysis}a shows the SNR as a function of integration time interval $T_M$ for various detector devices.  We used a device capacitance value of 1 pF for the UVPC, PMT and APD devices, and a per-pixel capacitance value of 0.1 pF for the CCD and EMCCD detectors (both values are dominated by stray capacitances). The SNR improves as the QE, gain and integration time increase, and as the ENF and dark current decrease. In a realistic state detection experiment, one needs  to distill the decision (for qubit state of 0 or 1) down to a single threshold parameter, either in terms of integrated current (analog) or detected photon number (digital), in a given measurement time interval $T_M$. Figure \ref{SNRAnalysis}b plots the optimal BER as a function of $T_M$ for these devices. The BER in general is determined by the SNR, but the ENF increases the BER due to the broadening of the signal arising from multiplication noise. High QE UVPC and EMCCD have the best promise in terms of low BER detection at minimal integration times. While EMCCDs feature attractive performance in terms of the QE, gain and EMF, the impact of intrinsic latency arising from the multiplication process on quantum error correction processes must be considered. This analysis shows that the minimum time required for a low BER state detection ($<10^{-3}$) is about $50 \mu$s. Further improvement in BER can come from higher photon scattering rates, enhanced collection efficiency and higher QE detectors. A potentially dramatic increase can be achieved if optical cavities with a small mode volume are used, as discussed in Section \ref{sec:Communication}.

\subsection{\label{subsec:Collection} Strategies for scalable and efficient photon collection}

Besides improving the detection efficiency, technologies that allow parallel photon detection operations also increase the system performance. Thousands of VLPCs have been used in high energy physics experiments with simple readout circuits \cite{AdamsIEEETNS1995}. Geiger mode APDs have a small detector area ($\sim 50 \mu$m in diameter) and high performance operations require sophisticated bias circuits, and might be more costly to scale to a large number of devices. The CCD and EMCCD cameras have the intrinsic benefit of being multi-pixel detectors. However, typical cameras have a relatively low frame rate ($\leq 100$Hz) due to a large number of pixels. By reducing the number of pixels (pixel binning), one can create high frame rate cameras that feature tens of thousands of frames per second \cite{SciMeasure}, although custom readout circuitry might be necessary for such a task. PMTs are typically bulky, and not suitable for scalable parallel operations.

High collection efficiency is achieved by using low $F/\#$ optics. If the collection optics are located outside the UHV chamber, the distance between the ion and the first optical element is quite large (several cm), leading to a bulky optical lens. Typically, a limited aperture is used in the imaging system that relays the scattered photons to the detector to limit the stray photons from reaching the detector, leading to a limited field of view \cite{KingThesis1999}. While this arrangement allows high fidelity measurements of qubits, it is not readily scalable to multiple locations and parallel operations.

In an ideal optical setup, one needs to locally magnify the ion detection regions, typically spread out over the ion trap chip, using low $F/\#$ optics. This cannot be achieved by using single field-of-view optics, indicating that multiple photon collection apertures are needed. One can utilize micro-optical components, like microlenses or lensed fiber tips, to achieve this goal by locally decreasing the $F/\#$. The focal length of a plano-convex lens is given by the index $n$ of the lens material and the radius of curvature $R$ of the convex surface by $f\simeq R/(n-1)$. Under this approximation, the $F/\#$  is given by $F/\# = f/D \simeq R/D(n-1)$, where $D$ is the diameter of the lens. One can see that $F/\# \simeq 1$ is difficult to achieve using conventional optical material with $n \leq 2$. When a lensed fiber is used, the effective $F/\#$ of the structure is given by $F/\# \simeq 1/2(NA_l+NA_f)$, where $NA_l$ and $NA_f$ are the numerical apertures of the lens and the fiber, respectively. This eases the requirement on $NA_l$ to be integrated at the tip. For example, when a multi-mode fiber with $NA_f = 0.22$ and diameter of 200 $\mu$m is used, one can achieve the overall $F/\# \simeq 1$ for the lensed fiber by creating a curved surface with a radius of curvature of ~220 $\mu$m at the fiber tip, if the average index of the fiber is about 1.48 (fused silica in the UV wavelength range).

 Figure \ref{DetectionStrategy}b shows an example where the tip of a fiber is modified to form a lens with a low $F/\#$ for effective photon collection. Using simple spherical surfaces ($R\sim 230 \mu$m), the $F/\#$ for a lensed fiber tip is estimated to be close to 1. Since the diameter of the fiber is typically less than 1 mm, this arrangement requires the fibers to be located very close to the ions through UHV fiber feedthroughs. Drawbacks of this approach include the challenge of creating low $F/\#$ lensed fiber tips, UHV-compatible fiber feedthroughs, and the electrostatic impact of glass fibers in close proximity to the trapped ions. A hybrid solution is shown in Fig. \ref{DetectionStrategy}c, where the scattered photons from ion traps are imaged onto an intermediate plane outside the vacuum chamber using a low $F/\#$, low magnification ($\sim$2-5) optics. Lensed fibers are aligned to the image of the ion locations, and only collect the scattered photons and direct them to the photon detectors. This scheme eliminates the need for locating the fiber very close to the ions and UHV fiber feedthroughs. The magnification substantially decreases the divergence angle of the scattered photons at the image plane, increasing the required $F/\#$ for the lensed fiber tips. The drawback is that the imaging lens with large $F/\#$ is still needed in this scheme, so locating the ion trap close to the UHV chamber window to reduce the volume of this lens is important. 
 
\section{\label{sec:Communication} Optical design for entanglement generation}
The entanglement between a single photon and a single ion was demonstrated experimentally using ultrafast excitation pulses \cite{BlinovNature2004}. Using two entangled photon-ion pairs, one can prepare two remote ions in an entangled state via entanglement swapping, which requires a joint measurement on the photons emitted from each ion. When the photons are combined at a beamsplitter and coincidence is detected, the two ions are left in an entangled state \cite{Maunz0608047,MoehringNature2007}. This procedure provides an exciting new paradigm for scaling an ion-trap based QIP, but faces the challenge that the success probability of the demonstrated entanglement generation is quite low ($10^{-8}-10^{-9}$). Such a low success probability in current experiments leads to long time constants for the entanglement generation process (1 event every 8.5 minutes), which can be improved by several orders of magnitude if the emitted single photon can be collected very efficiently. We also point out that the atomic states chosen to implement the entanglement generation protocol is not consistent with the choice of qubit states outlined in Section \ref{sec:PhysicalRequirements}, and a qubit translation operation is necessary to utilize the generated entanglement in local logical operations within the ELU.

\subsection{\label{MicroCavityDesign} Design considerations for on-chip micro-cavities}
Cavities constructed around the ion can substantially modify the spontaneous emission properties of the ions, and make the photon collection process dramatically more efficient \cite{BlinovNature2004,GuthohrleinNature2001,EschnerFdP2003,KrueterPRL2004}. In a recent experiment, a high quality factor (Q) cavity was constructed around a trapped ion and deterministic state evolution between the ion and a single cavity photon was demonstrated with an estimated single photon generation efficiency of $\sim 8\%$ and photon detection efficiency of $\sim 4.6 \%$ (overall yield of $\sim 0.4\%$ for detecting single photon when the ion is excited) \cite{KellerNature2004}. The coupling efficiency of the spontaneously emitted photon to a single cavity mode is dramatically enhanced when the cavity mode volume is reduced. Miniaturized versions of such cavities have been proposed in the context of single atom detection on a chip \cite{HorakPRA2003}, and simple prototype devices have been demonstrated \cite{TrupkeAPL2005,WilzbachFdP2006,SteinmetzAPL2006}. More recently, strong coupling of such micro-cavities and atoms has also been demonstrated \cite{ColombeNature2007}.

\begin{figure}[htbp]
	\centering
		\includegraphics[width=4in]{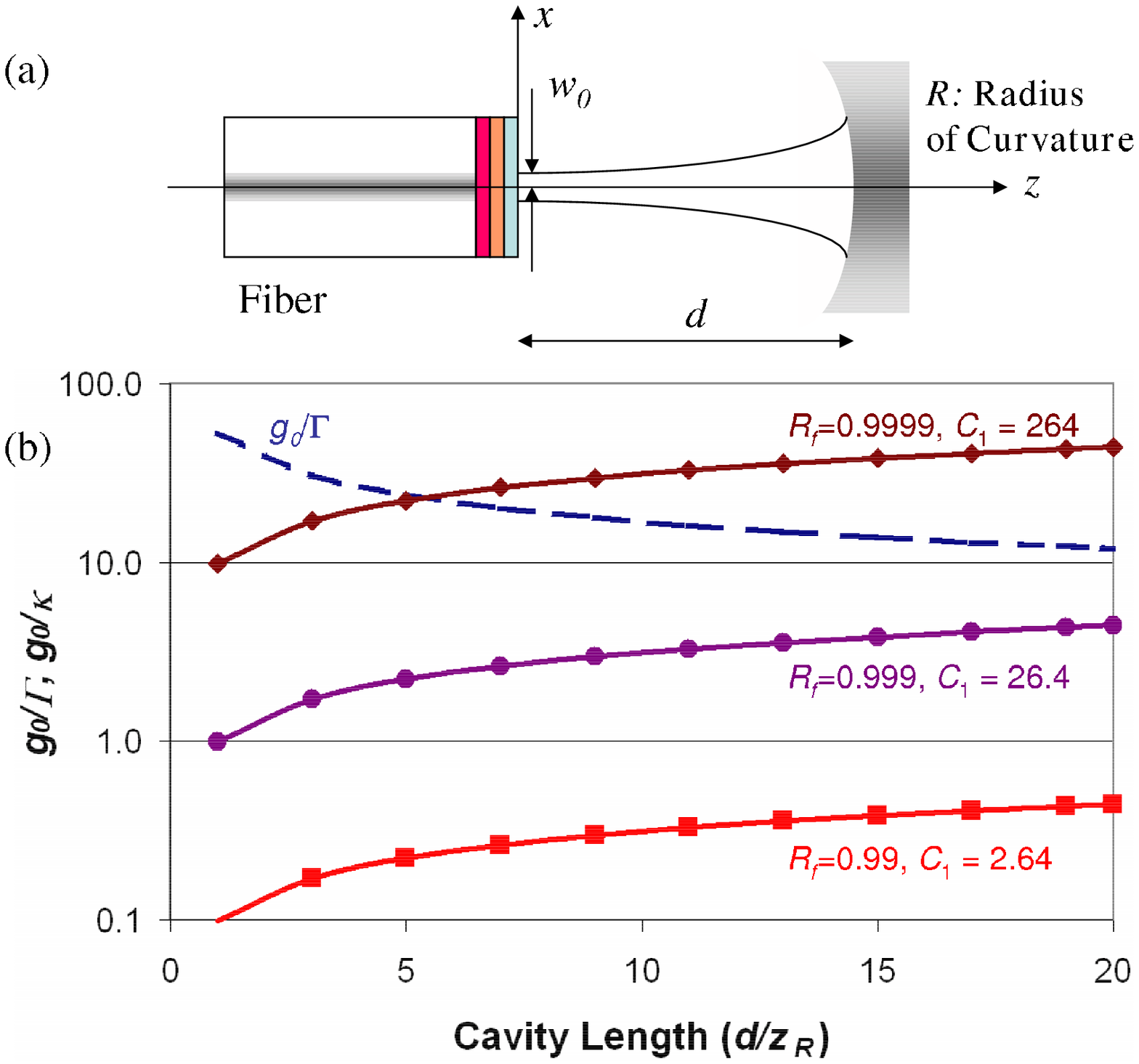}
		\fcaption{\label{GaussianMode}
(a) Schematic of the concave cavity constructed with a micro-fabricated concave mirror and a fiber tip. (b) The ratio of coherent coupling to dissipative mechanisms, $g_0 / \Gamma$ (dashed line) and $g_0 /\kappa$ (solid lines) as a function of the cavity length in units of the Rayleigh length $z_R$. This calculation is done for $^9$Be$^+$ ions placed in a cavity mode with a beam waist of $w_0 = 1.5 \mu$m and corresponding Rayleigh length $z_R = 22.6 \mu$m, assuming that the reflectance of the concave mirror $R_m=1$ so that the finesse $F$ of the cavity is determined by the reflectance $R_f$ of the fiber mirror. The ratio $g_0 / \Gamma$ is independent of the mirror reflectance. Three cases of mirror reflectances are shown, $R_f =$ 0.99 (square), 0.999 (circle) and 0.9999 (diamond). The cooperativity parameter is a unique function of the finesse once the beam waist and the wavelengths are fixed.}
\end{figure}

Here we consider a micro-cavity formed between a concave mirror with a radius of curvature $R$ and a tip of a single mode fiber (the diameter of the fiber core is assumed to be 3 $\mu m$) separated by a distance $d$ (Fig. \ref{GaussianMode}a). We assume anti-reflection (AR) coating with reflectances of $R_m$ and $R_f$ for the mirror and fiber tip, respectively. The reflectances determine the cavity finesse $F\simeq -\pi/ \ln (R_m R_f)$ and the cavity decay rate $\kappa = \pi c/dF$, where $c$ is the speed of light. In order to ensure that the cavity mode couples effectively into the fiber mode, we choose $1-R_m \ll 1-R_f$, so the output coupling is dominant through the fiber port. The scattering loss $L_s=1-\exp[-(4 \pi \sigma / \lambda)^2]$ due to surface roughness of the mirrors (characterized by the root-mean-square, or {\em rms}, roughness $\sigma$) can cause unwanted loss at the mirrors \cite{TrupkeAPL2005}. One needs to ensure that the loss due to this scattering is negligible compared to the transmittance of each mirror. This condition puts additional constraint on the quality of the mirrors, requiring that the {\em rms} roughness $\sigma < 0.8$nm ($0.25$nm) to reduce the scattering-related loss to below 0.1$\%$ (0.01$\%$) in the UV wavelengths of interest.

The spontaneous emission rate of the ion in free space is given by the Wigner-Weisskopf theory to be $\Gamma = \omega_a^3 p_d^2/ (3 \pi \epsilon_0 \hbar c^3)$, where $p_d$ and $\omega_a$ are the (atomic) dipole moment and the angular frequency of the optical transition under consideration, respectively, $\epsilon_0$ is the electrical permittivity of vacuum, and $\hbar$ is the Planck constant \cite{Louisell1973}. When the atom (or ion) is placed in a cavity, the cavity mode interaction with the atomic dipole can be described effectively using the Jaynes-Cummings model \cite{WallsMilburn1994}. The dynamics are described by the vacuum Rabi frequency between the atomic excitation and the cavity field $g_0 = \sqrt{2}p_d E/\hbar$, where $E$ is the electric field amplitude inside the cavity mode due to a single photon at the location of the ion.

A plano-concave cavity with a Gaussian mode beam waist of $w_0 = 1.5 \mu$m at the planar mirror can be created when the radius of curvature $R$ of the concave mirror and the cavity length $d$ satisfy the relationship

\begin{equation}
d(|R|-d)= z_R^2,
\label{GaussianCondition}
\end{equation}
where $z_R = \pi w_0^2/\lambda$ is the Rayleigh length of the mode and $\lambda$ is the transition wavelength \cite{SalehTeich1991a}. For cavities satisfying this condition, the mode volume is given by $V=\pi w_0^2 d$, and the electric field maximum due to a single photon of energy $\hbar \omega=hc/\lambda$ is given by

\begin{equation}
E_0 = \sqrt{\frac{4hc}{\lambda V \epsilon_0 (1+\lambda^2/2 \pi^2 w_0^2)}}=\sqrt{\frac{4hc}{\lambda V \epsilon_0 (1+\alpha)}},
\label{GaussianEmax}
\end{equation}
where $\alpha \equiv \lambda^2/2 \pi^2 w_0^2 \ll 1$ denotes the fraction of the energy contained in the electric field component parallel to the optical cavity axis. The electric field amplitude at an arbitrary location within the cavity mode can be calculated from this maximum field and the Gaussian mode function. Specifically, the electric field along the optical axis of the cavity ($x=y=0$) is given by
$|E(0,0,z)| = E_0/\sqrt{1+(z/z_R)^2}$, which allows us to calculate the vacuum Rabi frequency on the optical axis of the cavity. For efficient coupling of the spontaneously emitted photon into the fiber mode, we require that the atomic excitation coherently transfers the excitation to the cavity mode at a rate much faster than the spontaneous emission rate ($g_0 \gg \gamma$), and that the excitation in the cavity leak out into the fiber mode much faster than it can be reabsorbed by the atom ($g_0 \ll \kappa$). The cooperativity parameter is defined by 

\begin{equation}
C_1 \equiv g_0^2/ 2 \Gamma \kappa = \frac{3 \lambda^2 F}{\pi^3 w_0^2 (1+\alpha)},
\label{CooperativityParam}
\end{equation}
and is uniquely determined by the finesse of the cavity when the wavelength and the beam waist of the mode are defined. The fraction $f_{\mathrm{cap}}$ of photons emitted into the fiber mode from the atom is given by the cooperativity parameter as $f_{\mathrm{cap}}=2C_1/(2C_1+1)$. With $C_1 \sim 2 (20)$, one can capture $\sim 80\%$ ($\sim 98 \%$) of the photons spontaneously emitted by the photon into the fiber mode. Since the success rate of the entangling scheme is proportional to $f_{\mathrm{cap}}^2$, one can improve the generation rate of the entangled pairs by a factor of $4 \times 10^4$ if $f_{\mathrm{cap}}$ is increased from $0.4\%$ to $80 \%$.

Figure \ref{GaussianMode}b shows the ratio of the Rabi frequency to the dissipative mechanisms in the system, $g_0 /\Gamma$ (dashed line) and $g_0 /\kappa$ (solid lines), as a function of the cavity length (measured in the units of Rayleigh length, $z_R = 22.6 \mu$m in this case) as the finesse of the cavity is varied. We assume the concave mirror has a unity reflectance, and the cavity finesse is determined by the reflectance $R_f$ of the fiber mirror. The vacuum Rabi frequency and the spontaneous emission rate do not depend on the finesse of the cavity, and we see that for small cavity mode volumes, this ratio can be kept much larger than unity. As the reflectance of the cavity increases from 0.99 (square) to 0.999 (circle) to 0.9999 (diamond), the ratio $g_0 /\kappa$ increases. The system enters a strong coupling regime ($g_0 \gg \Gamma, \;\; \kappa$) when the mirror reflectance is above 0.999. It is important to note that the atomic excitation can be efficiently collected as a photon in the fiber without necessarily being in the strong coupling regime, with a modest mirror reflectance when the cavity mode volume is made small. A similar cavity with a multi-mode fiber on one side can be used for high-speed state detection by increasing the collection efficiency of the photons scattered during the cycling transition.

It is important to note that the presence of the fiber and the curved mirror at the vicinity of the ion will significantly alter the trapping mechanism of the Paul trap. In order to realize this system, an innovative trap design is necessary to integrate the trapping electrodes with the cavity geometry.

\section{\label{sec:Conclusions} Conclusions}
In this paper, we have considered the requirements of constructing a scalable ion trap QIP, and explored the system design strategies and critical components necessary for achieving the integration of the optical functions in the QIP. Realization of the optical elements outlined here provides the technology platform for constructing a large scale QIP based on a quantum network of integrated trapped ion chips.

\nonumsection{Acknowledgements}
\noindent
The authors would like to thank the David Wineland, Roee Ozeri, Diedrich Leibfried and the rest of the Ion Storage Group at NIST in Boulder, Chris Monroe and Dick Slusher for helpful discussions and ideas. This work was supported by NSF under CCF-0520702 and CCF-0546068.

\nonumsection{References}

\end{document}